# A New Underground Laboratory in the USA for a Neutrino Factory Detector and Other Scientific Projects*

David B. Cline

*Center for Advanced Accelerators*
*Department of Physics and Astronomy, Box 951547*
*University of California, Los Angeles, CA 90095-1547 USA*

**Abstract.** A neutrino factory storage ring can provide beams to various locations. We discuss the ICANOE detector (at LNGS) at one such site. We then describe the prospects for the use of the underground location at Carlsbad, NM for a neutrino factory detector. A brief discussion is given of a simple magnetized Fe detector of 10 – 50 kT for this site.

## INTRODUCTION

The development of a neutrino factory requires the complimentary development of a large detector, possibly 50 kT. It is likely that such a detector will go underground. There are actually only a few underground laboratories in the world capable of hosting such a detector and very few in the USA. Of course the LNGS is a wonderful example of a laboratory with a large overburden that is on the surface, and the ICANOE detector is possibly useful for a neutrino factory if the magnetized Fe system is rearranged. In this report, we describe a new underground laboratory for particle physics application and the possible development of a neutrino factory detector for this site at Carlsbad, NM.

## PHYSICS POTENTIAL OF THE NEUTRINO FACTORY

The physics potential of a muon storage-ring neutrino factory is large[1-3] and includes

(A) Observation of $\nu_e \to \nu_\mu$ at levels of $10^{-3}$; possible detection of $\nu_e \to \nu_\mu$ and $\nu_e \to \nu_\tau$;
(B) Study of $\nu_\mu \to \nu_\tau$ and $\bar{\nu}_\mu \to \bar{\nu}_\tau$;
(C) Possible CP violation test for
  (i) $\nu_\mu \to \nu_\tau$, $(\bar{\nu}_\mu \to \bar{\nu}_\tau)$;
  (ii) $\nu_e \to \nu_\tau$, $(\bar{\nu}_e \to \bar{\nu}_\tau)$, $\nu_e \to \nu_\mu$, $(\bar{\nu}_e \to \bar{\nu}_\mu)$;
(D) Possible study of the combined effects $\nu_\mu, \bar{\nu}_\mu$ to $\nu_e, \bar{\nu}_e$ final states (T violation).

In the case (Ci), we expect a very small signal in the three-neutrino "CKM" matrix formulation. However, if there are sterile neutrinos, the CP violation could still be very large in this channel.

One strength of the ICANOE[4] detector is the real-time identification of

$$\nu_\tau + N \rightarrow \tau + x$$
$$\quad \hookrightarrow e,\mu$$
$$\quad \hookrightarrow \text{hadron} \quad .$$

Using NOMAD-like cuts in addition, the process $\nu_e \rightarrow \nu_\mu$, $\nu_\mu + N \rightarrow \mu^- + x$ can be detected using a magnetized toroid system such as is employed for ICANOE.

In order to address the detection of the processes described before and the resulting background, we need to specify the neutrino factory parameters.

We point out for historical reference that one of the earliest papers on a $\mu^\pm$ neutrino source was by Cline and Neuffer in 1979-80.[1] The recent interest in this subject is due to the probable discovery of neutrino oscillations by Superkamiokande (SK) using atmospheric neutrinos.

The current situation in neutrino mass and mixing physics lends itself to some confusion. We do not know if there are sterile neutrinos or not, and a neutrino factory could be the key to resolving this issue.

Another key possibly is to operate a neutrino factory below $\nu_e \rightarrow \nu_\tau$ threshold and then observe $\bar{\nu}_e \rightarrow \bar{\nu}_\mu \rightarrow \mu^-$ in the detector. The detection of $\nu_e \rightarrow \nu_\tau$ or $\nu_e \rightarrow \nu_\mu$ will be crucial for the use of a neutrino factory to study the neutrino CKM matrix and to search for CP or T violation.

## Backgrounds to Neutrino Oscillation Measurements

The basic concept of the neutrino factory is to detect an oscillation in the non-$\nu_\mu$ channel. For example for a $\bar{\mu}$ storage ring, $\bar{\mu} \rightarrow \bar{e} + \nu_\mu + \bar{\nu}_e$. So we have $\bar{\nu}_e$ in the $\nu_\mu$ beam, oscillation into $\bar{\nu}_e \rightarrow \bar{\nu}_\mu$, followed by $\bar{\nu}_\mu \rightarrow \mu^+$ given a "wrong sign lepton." In the case of $\nu_e \rightarrow \nu_\tau$, as we will show, the produced $\tau$ must decay into a $\mu^\pm$ channel to measure the lepton charge – the so-called wrong-sign lepton.

We now discuss the possible level of the signals for neutrino oscillations at a neutrino factory. We can write the transition probably as

$$P_{\mu\tau} = \cos^2\theta_{13}\sin^2 2\theta_{23}\sin^2\left[1.27\left(\frac{L}{E}\right)\Delta^2 m\right] \quad ,$$

$$P_{e\tau} = \sin^2\theta_{23}\sin^2 2\theta_{13}\sin^2\left[1.27\left(\frac{L}{E}\right)\Delta^2 m\right] \quad ,$$

where L is in km, E is in GeV, and $\Delta^2 m$ is in $eV^2$. The SuperKamiokande results suggest that $\theta_{23} \approx 45°$. The current limits on $\nu_\mu \rightarrow \nu_e$ constrain $\theta_{13}$ to be less than 20°. Since we know nothing about the neutrino CKM matrix, $\theta_{13}$ could be very small. We can express the $\nu_e \rightarrow \nu_\tau$ transition as $P_{e\tau} \approx P_{e\mu} (\cos^2\theta_{23}/\sin^2\theta_{23})$. For $\theta_{23} \simeq 45°$, we find $P_{e\tau} \approx P_{e\mu}$). Thus for the issue of background estimates, if $\theta_{13}$ is small, then both $P_{e\tau}$ and $P_{e\mu}$ will be small.

There are natural sources of wrong sign leptons produced in high-energy neutrino interactions. While these backgrounds will not likely affect the detection of $\nu_e \rightarrow \nu_\mu$, we will show that for some parameters this background will be dangerous for $\nu_e \rightarrow \nu_\tau$ detection.

To show why wrong-sign leptons produced in neutrino interactions may be a limiting background for $\nu_e \to \nu_\tau$, we consider the following:

$\nu_e \to \nu_\tau \quad , \quad (P_{e\tau}) \quad ;$

$\nu_\tau + N \to \tau^- + X \quad , \quad (\sigma_\tau/\sigma_{all}) \quad ;$

$\tau \to \mu^- + X \quad , \quad (P \sim 10^{-1}) \quad .$

(Note that the $\mu^-$ energy spectrum will be soft.)

The $\mu^\pm$ must be detected to identify the "wrong sign $\tau$" from the large rate of production from $\bar{\nu}_\mu \to \bar{\nu}_\tau$ in the neutrino factory beam. For the parameters $P_{e\tau} \sim 5 \times 10^{-3}$, $\sigma_\tau/\sigma_{all} \sim 5 \times 10^{-2}$. We find the overall probability to get a wrong sign $\mu^\pm$ from an ($e\tau$) oscillation to be $2.5 \times 10^{-5}$.[5,6] We will show that this is below some of the "natural" wrong sign production in neutrino interactions.

The production of charm can lead to wrong sign leptons through the process

$\nu_\mu + N \to \mu^- + \text{charm}$

$\qquad\qquad\qquad\qquad \hookrightarrow \mu^+ + \nu + X \quad ,$

with the $\mu^-$ being missed. (This process was first observed by the HPWF group at FNAL in 1974.) We do not expect this to be an important background for $\nu_e \to \nu_\mu$ because of the high energy of the resulting $\mu^-$. However, for $\nu_e \to \nu_\tau$, the $\tau$ decay into a $\mu^-$ channel will result in a soft $\mu^-$, as well as some missing neutrino energy and the same hadronic structure as charm production.

## DETECTORS AT THE GRAN SASSO AND CARLSBAD SITES

We assume that distances of 2000 to 7000 km are useful to study neutrino oscillations, matter effects, and possibly CP violation. We also assume that the detectors must be located underground, because of the enormous backgrounds from cosmic rays. Finally, we assume that existing underground facilities will be used for the detector. In this section, we discuss detector locations of ~ 2000 to 3000 km from either FNAL or BNL and 6000 to 7000 km to the LNGS.

We discuss two detectors:

a. ICANOE (6-kT of liquid argon) at the LNGS [Fig. 1(A) and (B)] and
b. A magnetized Fe detector at the Carlsbad Underground Laboratory (Fig. 2 and Tables 1 and 2).

These two detectors are in a sense the extremes of the detectors one may use for a neutrino factory. The ICANOE detector will have excellent electron identification and can observe $\nu_{\mu,\tau} \to \nu_e$, whereas the Carlsbad detector will likely be used for wrong-sign muon identification, i.e., $\bar{\nu}_e \to \bar{\nu}_\mu \to \mu^+$ in the detector. To search for CP violation, both detectors will have to use $\mu^\pm$ sign determination, since we have to way to measure the electric charge of the "*e*"-like signals.

We will first address the issue of the crossing angle of the neutrino beam on the detector from a U.S. based (BNL or FNAL) neutrino factory . This will not be a problem for a new hall as can be constructed at the Carlsbad site, but as shown in Fig. 1(B), it could be a problem for a fixed laboratory like the LNGS. It is possible that part of the ICANOE detector can be reconfigured to fit this space angle.

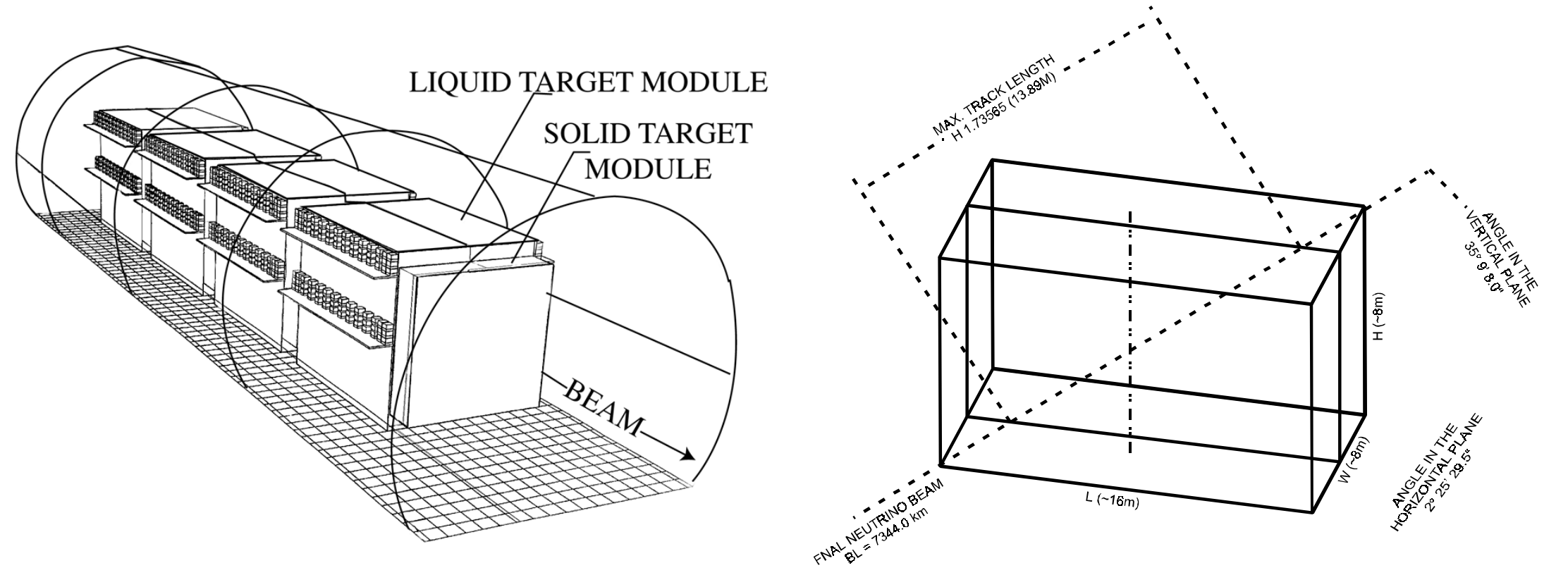

(A) (B)

**FIGURE 1.** (A) Schematic of the ICANOE detector to be constructed at the LNGS; (B) Detector crossing of the LBL beam from FNAL for the ICANOE detector at LNGS.

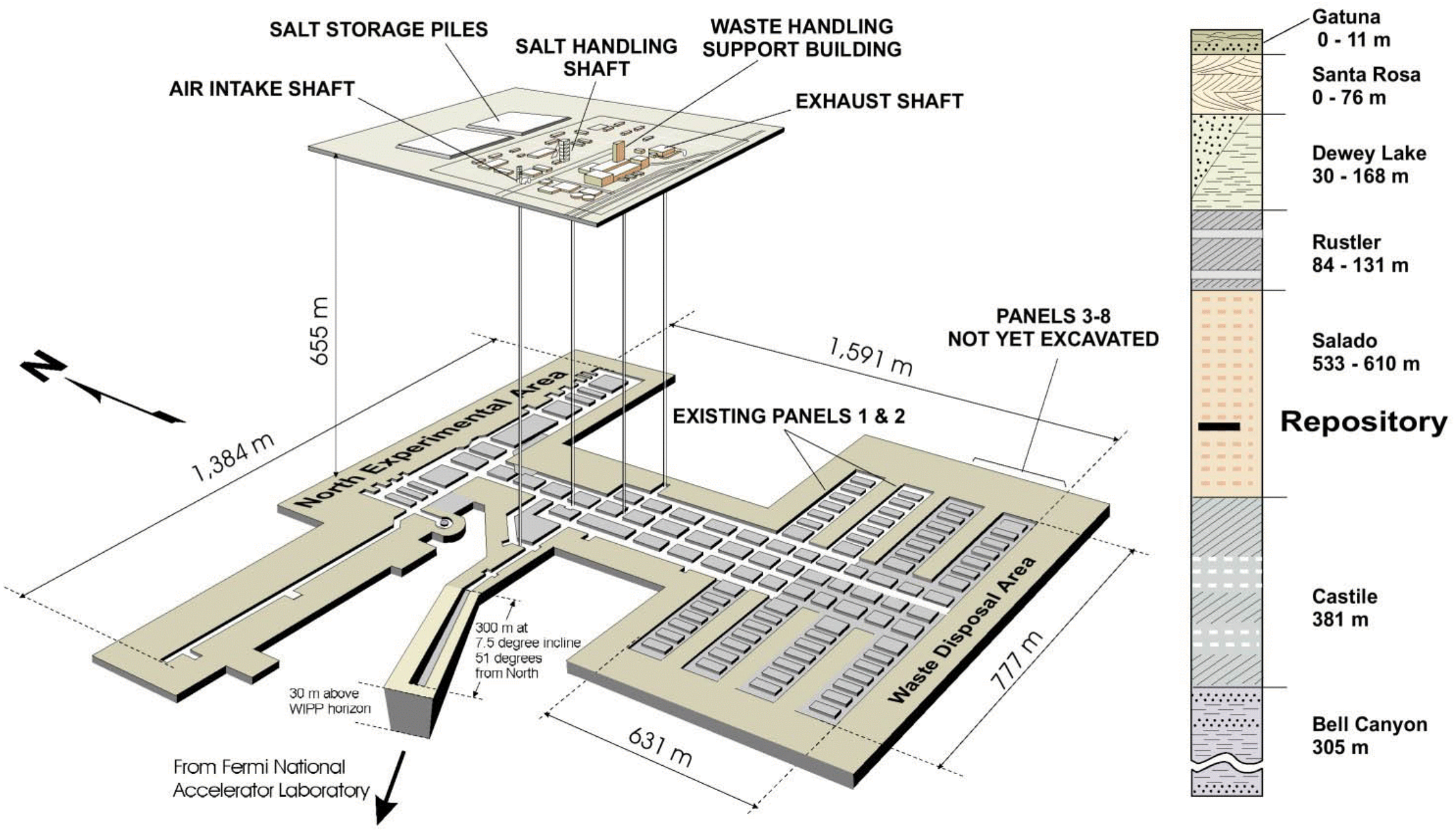

**FIGURE 2.** Possible neutrino-factory detector at the WIPP site in Carlsbad, NM.

## DEVELOPMENT OF THE UNDERGROUND LABORATORY AT CARLSBAD, NM

In order to establish a new underground laboratory for neutrino physics, other projects besides a neutrino factory detector must be developed. A powerful elevator is required to construct a large detector. As can be seen from Tables 1 and 2, such an elevator exists at the Carlsbad site.

The site also has a very low radioactive background (Table 3). This makes it suitable for other experiments (e.g., WIMP searches, ββ decay search, a novel supernova detector, etc.). The depth of the site also helps (see Fig. 3). As can be seen from the figure, the underground site is in halite (or mainly NaCl structure), which provides for easy excavation for the tunnel for the neutrino factory detector (see Fig. 2). Also the halite has low radioactive backgrounds, which are needed for other projects at such a site.

## A NEUTRINO FACTORY DETECTOR AT CARLSBAD

In a related work,[8] we have described a bowtie storage ring to send beams to several detectors. There are two possible detectors that could be installed at the Carlsbad to use for a neutrino factory:

1. A large ICANOE-type detector (as I noted in Ref. 6, which was presented at the first neutrino factory meeting in Lyon), and
2. A large, relatively inexpensve magnetized Fe spectrometer.[9]

In Fig. 4, we show a schematic of a 10-kT magnetized Fe detector. Such a detector could also be scaled to 50 kT. While Fig. 2 shows a tunnel that points to FNAL, it is also possible to point the tunnel in the direction of BNL or even to CERN if that is where a neutrino factory is eventually constructed.

## ACKNOWLEDGMENT

We wish to thank Y. Fukui, A. Garren, and F. Sergiampietri for their help. We also thank R. Nelson for the use of WIPP site drawings (Figs. 2 – 4).

Table 1. Carlsbad Site[7]

- Federal facility with > 35-yr lifetime
- 661-m below surface (2000 mwe*)
- Extensive facility infrastructure in place:
  - U/G transport and personnel/equipment access
  - Power and high-speed data communications
  - Security, safety, and emergency services
  - Surface support and highway access

*Meters water equivalent

Table 2. WIPP Underground Research Facility[7]

- 40-ton material/personnel shaft
  - 9'5" wide × 15'4" deep × 13'8" high (2 cages)
  - Nominal 15 trips/day
- 20-ton salt shaft provides added flexibility (plus 2 ventilation shafts)
  - 3'11" wide × 4'3" deep × 11'8" high (2 cages)
- Two 13.8-kV circuits with 60 A/480 V/3-phase
- Twenty-two fiber optic lines with Tbase 10 (72 planned by 2002)

Table 3. Summary of Background Radiation at WIPP[7]

| | | |
|---|---|---|
| • Thorium (ppb) | 96 ± 59 | (27 – 195) |
| • Uranium (ppb) | 49 ± 24 | (19 – 91) |
| • Neutron Flux | | |
| - | Total: ~340 $m^{-2}$ $d^{-1}$ | |
| - | Epithermal and thermal: ~115 $m^{-2}$ $d^{-1}$ | |

- Radon from surface air dominates levels in underground
- Depth ~2000 mwe (close to Soudan): $\mu \sim 1-2 \times 10^{-7}$ $cm^{-2}$ $s^{-1}$
- No significant increase (from WIPP TRU waste) in neutrino environment above that from primordial terrestrial sources (K/Th/U)

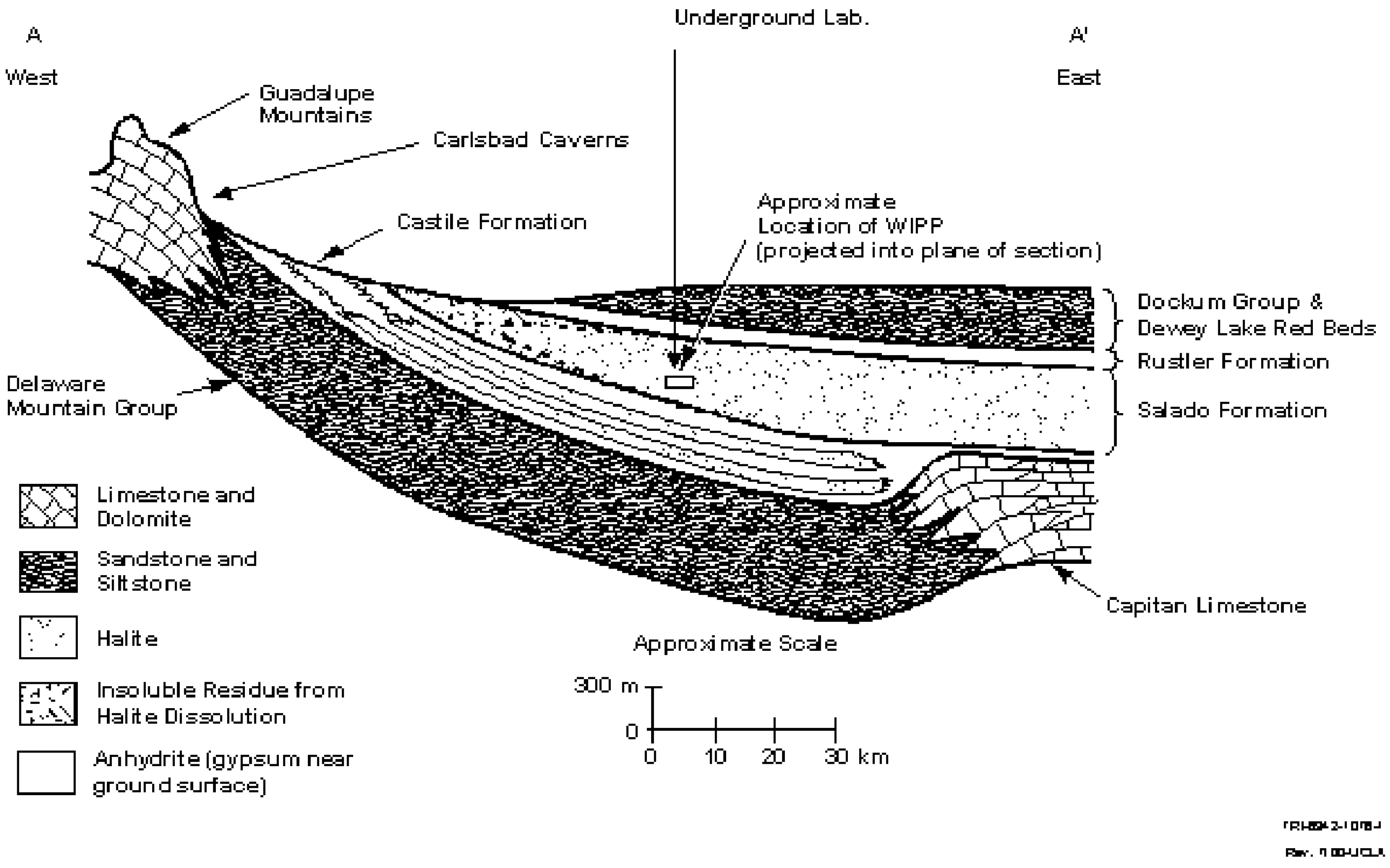

Fig. 3. Cross section of the underground structure near the Carlsbad site.

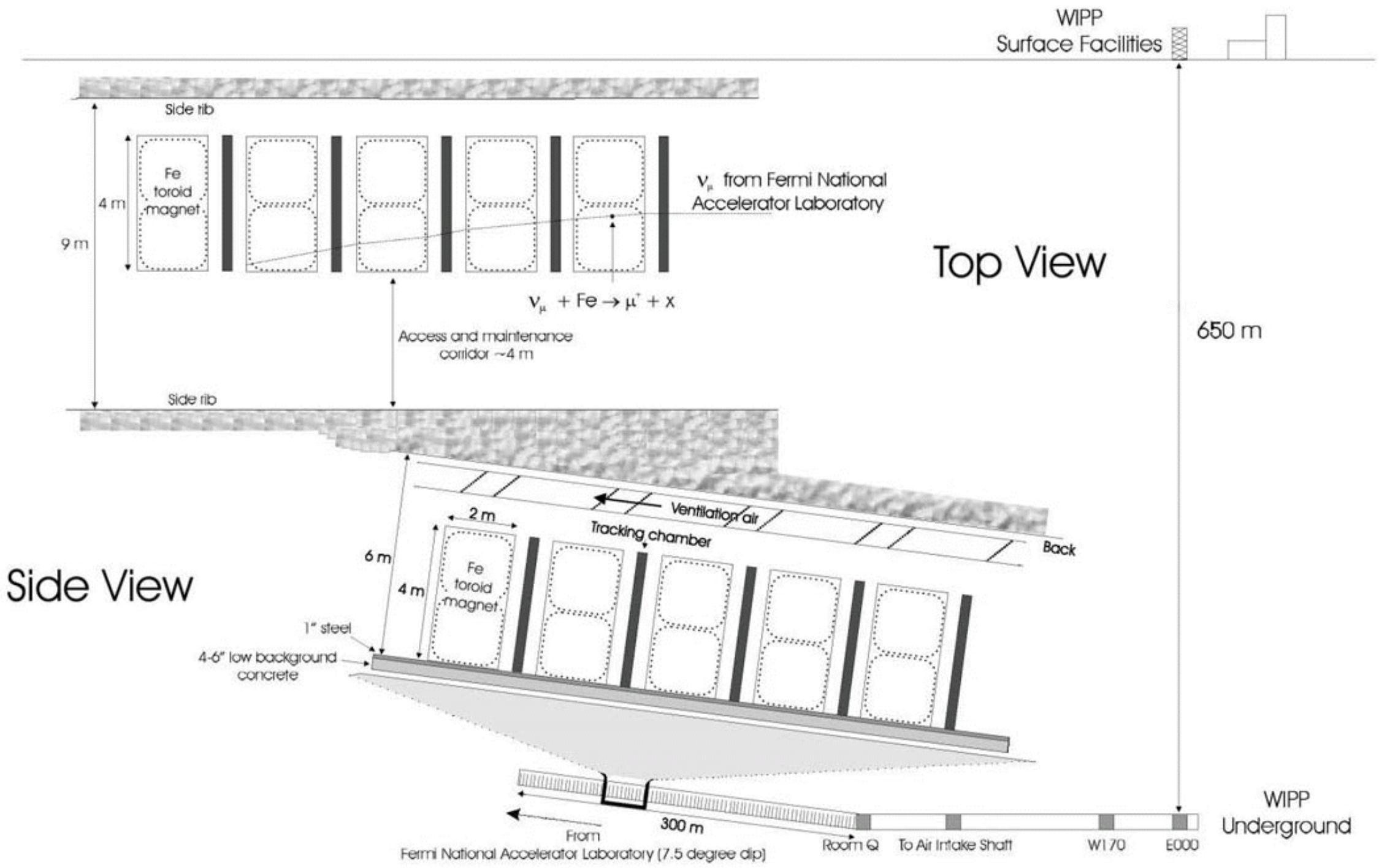

Fig. 4. Schematic of a 10-kT, magnetized Fe spectrometer for a neutrino factory detector at Carlsbad.